\documentclass[showpacs,preprintnumbers,amsmath,amssymb]{revtex4}
\usepackage{cases}
\usepackage{amsmath}
\usepackage{amssymb}
\usepackage{amsfonts}
\usepackage{amssymb}
\usepackage{dcolumn}
\usepackage{bm}
\usepackage{epsfig}
\usepackage{bbm}
\usepackage{graphicx}
\usepackage{xcolor}
\usepackage{array}
\usepackage{subfigure}
\usepackage{hyperref}
\usepackage{mathrsfs}
\usepackage{verbatim}
\usepackage{float}
\usepackage{epstopdf}
\usepackage{CJKutf8}
\def\bit{\begin{itemize}}
\def\eit{\end{itemize}}
\def\ben{\begin{enumerate}}
\def\een{\end{enumerate}}
\def\bed{\begin{description}}
\def\eed{\end{description}}

\def\lsim{\raise0.3ex\hbox{$<$\kern-0.75em\raise-1.1ex\hbox{$\sim$}}}
\def\gsim{\raise0.3ex\hbox{$>$\kern-0.75em\raise-1.1ex\hbox{$\sim$}}}

\let\jnfont=\rm
\def\NPB#1,{{\jnfont Nucl.\ Phys.\ B }{\bf #1},}
\def\PLB#1,{{\jnfont Phys.\ Lett.\ B }{\bf #1},}
\def\EPJC#1,{{\jnfont Eur.\ Phys.\ Jour.\ C }{\bf #1},}
\def\PRD#1,{{\jnfont Phys.\ Rev.\ D }{\bf #1},}
\def\PRL#1,{{\jnfont Phys.\ Rev.\ Lett.\ }{\bf #1},}
\def\MPLA#1,{{\jnfont Mod.\ Phys.\ Lett.\ A }{\bf #1},}
\def\JPG#1,{{\jnfont J.\ Phys.\ G}{\bf #1},}
\def\CTP#1,{{\jnfont Commun.\ Theor.\ Phys.\ }{\bf #1},}
\def\JHEP#1,{{\jnfont JHEP \ }{\bf #1},}
\def\NPPS#1,{{\jnfont Nucl.\ Phys.\ Proc.\ Suppl.\ }{\bf #1},}

\def\beq{\begin{equation}}
\def\eeq{\end{equation}}
\def\bea{\begin{eqnarray}}
\def\eea{\end{eqnarray}}
\newcommand{\ba}{\begin{array}}
\newcommand{\ea}{\end{array}}

\usepackage{color}

\begin{document}
\title{The most probable distributions
with finite number of particles and 
the applications on Bose-Einstein condensation}
\author{Qi-Wei Liang}
\email{qiweiliang@emails.bjut.edu.cn}
\affiliation{Faculty of Science, Beijing University of Technology, Beijing, China}
\author{Wenyu Wang}
\email{wywang@bjut.edu.cn}
\affiliation{Faculty of Science, Beijing University of Technology, Beijing, China}

\begin{abstract}
Motivated by the Asynchronous Finite Differences Method utilized for the calculation of the most probable distributions of finite particle number systems, this study employs numerical variation and central difference techniques to provide more precise estimations regarding these distributions. Specifically, three novel finite distributions are derived and applied to Bose-Einstein condensation, revealing that the critical condition ($n\lambda^3=2.612$) may be relaxed in finite particle number scenarios. Moreover, maintaining density as a constant is anticipated to result in a higher critical temperature compared to infinite number systems. Notably, the obtained condensate number on the zero-energy level surpasses that of predictions generated by canonical distributions.
\end{abstract}
\maketitle

\section{Introduction}\label{sec1}

The main objective of statistical physics is to comprehend and predict the macroscopic phenomena exhibited by systems comprising of an enormous number of microscopic particles, typically quantified by the Avogadro's constant ($6.022\times 10^{23}$). In statistical analysis, the infinity of micro-particles is considered as a fundamental concept for both equilibrium and non-equilibrium states. The abundance of micro-particles necessitates the transformation of functions defined on a discrete lattice into continuous ones, enabling the conduct of differentials, derivatives, and integration of such continuous functions during analysis. This concept is well-founded in statistical physics. For instance, the Stirling approximation of the factorials is a valuable computational tool in the determination of the most probable distribution (MPD) exhibited by systems such as Maxwell-Boltzmann, Bose-Einstein and Fermi-Dirac distributions. However, the application of infinite number postulation may pose a challenge while dealing with nano-scale physics,
levitation~\cite{gonzalez2021levitodynamics,jain2016direct,2018Levitated},
cold atom~\cite{andrews1997observation} or Bose-Einstein
condensate~\cite{anderson1995observation,davis1995bose,
bradley1995evidence}{\it etc.} 
Using the example of a levitated nanoparticle in Ref.~\cite{jain2016direct}, it is noted that the number of molecules present is of the order of approximately $10^{5}$. In view of this, treating the quantity as infinite appears to be an inaccurate approach. It is therefore recommended to analyze the quantity in terms of finite numbers, resulting in discrete functions. Derivatives, differentials and integration can then be converted to differences, ratios and summations. This allows for a more insightful comparison between the statistics obtained from an infinite number of particles versus a finite number of particles. This difference is of considerable interest to researchers.

The present study introduces a novel approach to perform statistical analyses on limited particle counts, as described in Ref.~\cite{Liu_2022}. A new variation of discrete calculus is presented, which employs asynchronous usage of forward and backward differences for the calculation of the Bose-Einstein, Fermi-Dirac, and Maxwell-Boltzmann systems. While the MPD functions conform to the established form when the number of particles exceeds one, the functions exhibit disparities when only one particle is present at a given energy level, rendering it indistinguishable as a Boson or Fermion. Despite its proficiency, the methodology is not without its drawbacks, such as the introduction of artificiality through the asynchronous application of forward and backward differences on distinct terms of a distribution, as highlighted in the subsequent discussion
(see the following context). 
In reality, the mathematical operations of integration and differentiation are founded upon the constrictions of discrete summation and difference. When the overall count of particles is meager, it is quite probable that the aforementioned operations will revert to their discrete configurations. Nonetheless, while discrete summation and difference hold the status of elementary approximation algorithms, they remain indispensable techniques for conducting numerical calculations. The successful implementation of these algorithms can furnish informative cues aimed at refining the precision of computations performed on finite number systems.

In this paper, inspired by the discrete numerical method, we have identified a more appropriate approach for dealing with finite number systems, numerical variation, and central difference methods. Our method provides more accurate variations for finite number systems. When we applied these new distributions to Bose-Einstein condensation, we observed many interesting features.
Overall, the contents of this paper are organized as follows. In sec.~\ref{sec2}, we introduce and analyze the method proposed in Ref. \cite{Liu_2022}. We also present the numerical methods in calculus that can be used for our analysis. In sec.~\ref{sec3}, we demonstrate the MPD using a more precise numerical calculus and study the implications of Bose-Einstein condensation. Finally, we present our conclusions in sec.~\ref{sec5}.

\section{Asynchronous finite differences and  hints from 
numerical variations and  differences}\label{sec2}

\subsection{Asynchronous finite differences}\label{liu-intro}

In order to compute the variations of finite discrete distribution functions, we adopted an approach referred to as asynchronous differences as proposed in Ref.~\cite{Liu_2022}. We briefly elaborate on this method below.
Numerical differentiation is commonly employed to approximate the value of a function's derivative at a specific point through a linear combination of its function values. A satisfactory formulation of discrete calculus of variation must account for both synchronous finite differences and asynchronous finite differences. This enables the finite differences-based solution groups, which consider all feasible combinations of forward and backward differences.
The definition of the derivative can be effectively utilized to employ the difference quotient for derivative approximation. Given the various directions of steps and the various
selections of approximate points chosen for calculating the slope,
numerous numerical differentiation formulas of a
function $y(x)$ can be obtained
\begin{eqnarray}
y_f^{\prime}(x)&\simeq&\frac{y(x+h)-y(x)}{h}\, , \label{eq:1a}  \\
y_b^{\prime}(x)&\simeq&\frac{y(x)-y(x-h)}{h}\, , \label{eq:1b} \\
y_c^{\prime}(x)&\simeq&\frac{y(x+h)-y(x-h)}{2h}\, , \label{eq:1c}
\end{eqnarray}
where $h$ is the step size increment. Note that
here the symbol ``$\simeq$" is chosen to represent 
``$=$" for a finite $h$. 
The equality will become exact when $h$ is a infinitesimal.
Equation (\ref{eq:1a}) is refereed to the forward difference quotient,
equation (\ref{eq:1b}) is refereed to the backward difference quotient,
and equation (\ref{eq:1c}) is the central difference quotient.
The forward and backward differences were adopted, 
i.e. the forward difference 
\begin{eqnarray}
\triangle_fy(x)=y(x+h)-y(x)\,,
\end{eqnarray}
 and the backward difference
 \begin{eqnarray}
 \triangle_by(x)=y(x)-y(x-h)\,.
 \end{eqnarray}
Doing these two differences on the different parts of 
a function will give asynchronous  differences. Take
\begin{equation}
    y(x)=y_{1}(x)+y_{2}(x)
\end{equation}
as an example, the differences 
$\triangle y(x)$ will have four combinations
\begin{eqnarray}
&& \triangle_fy_{1}(x)+\triangle_fy_{2}(x)\label{d1ff}\, , \\
&& \triangle_fy_{1}(x)+\triangle_by_{2}(x)\label{d1fb}\, ,  \\
&& \triangle_by_{1}(x)+\triangle_fy_{2}(x)\label{d1bf}\, ,  \\
&& \triangle_by_{1}(x)+\triangle_by_{2}(x)\label{d1bb}\, .
\end{eqnarray}
Differences Eq.~\eqref{d1fb} and Eq.~\eqref{d1bf} are the defined
asynchronous differences. The second order differences 
are also taken to one-to-one correspondence
with the first order differences 
\begin{eqnarray}
&&\triangle_f^{2}y_{1}(x)+\triangle_f^{2}y_{2}(x)
\label{d2ff}\,, \\
&&\triangle_f^{2}y_{1}(x)+\triangle_b^{2}y_{2}(x)
\label{d2fb}\,, \\
&&\triangle_b^{2}y_{1}(x)+\triangle_f^{2}y_{2}(x)
\label{d2bf}\,, \\
&&\triangle_b^{2}y_{1}(x)+\triangle_b^{2}y_{2}(x)
\label{d2bb}\,. 
\end{eqnarray}

By these asynchronous differences, the most probable distribution
can be derived by the variation on a microstates of a system with finite
number particles. Taking the Bose-Einstein distribution as an example,
the number of the distinct microstates in a set $\{n_i \}$ is 
\begin{eqnarray}
\Omega_{\rm B. E.}\{n_{i}\}=\prod_{i}\frac{(n_{i}+g_{i}-1)!}{n_{i}!(g_{i}-1)!}\,,
\end{eqnarray}
where $g_i$ is the degree of degeneracy of the energy level 
$\varepsilon_i$.
The functional $y$  with two Lagrange multipliers $\alpha$
and $\beta$ is
\begin{eqnarray}
y=\sum_{i}(\ln(n_{i}+g_{i}-1)!-\ln n_{i}!-\ln(g_{i}-1)!)-\alpha(\sum_{i}n_{i}-N)
-\beta(\sum_{i}n_{i}\varepsilon_{i}-E)\,,
\end{eqnarray} 
where  $N$ is the total 
particle number and $E$ is the total energy of the system.
The most probable distribution is derived by the variational 
\begin{eqnarray}
\delta y=\sum_{i}\delta n_{i}\left[\left(\frac{\delta\ln(n_{i}+g_{i}-1)!}{\delta n_{i}}-\frac{\delta\ln n_{i}!}{\delta n_{i}}\right)-(\alpha+\beta\varepsilon_{i})\right]=0\,.  \label{eq:v}
\end{eqnarray}
Here, we use the Stirling approximation to derive the MPD for a system with an infinite number of particles ($n_i \gg 1$). However, this approximation is invalid for finite numbers, and the asynchronous difference method is a viable solution.
The term $(\alpha+\beta\varepsilon_{i})$ is linear, and the asynchronous difference can be applied to the other two terms, denoted as $\ln(n_{i}+g_{i}-1)!$ (represented by $y_1$) and $\ln n_{i}!$ (represented by $y_2$).
Different asynchronous differences result in different distributions. The MPD is derived from the largest second-order variations $\delta^2 y$, which are described in Eq.\eqref{d2fb}, 1f2b for $n_i\ge 2$. This leads to the ordinary Bose-Einstein distribution as the final result.
For energy levels with a maximum of one occupied particle, the exact Bose-Einstein distribution can be derived using asynchronous differences, as detailed in Ref.\cite{Liu_2022}. Similar results can also be obtained for the other two distributions.

Although the use of the Stirling approximation has been abandoned, and the asynchronous difference method appears to have solved the MPD issue in finite particle number systems, there are still some defects that should be addressed.
First, the distribution function is artificially divided into two parts and combined with different differences, potentially compromising the stability of the algorithm. Upon analyzing relevant numerical data, it is unclear whether this method is stable.
Secondly, the true solution is based on the size of the second order difference, but the assumption of negative infinity is not acceptable when dealing with finite difference in discrete systems.
Lastly, the distribution functions are forced to be integer numbers if the continuous approximation is not used. However, the transition from discrete summation to continuous integration is the fundamental idea of calculus. Abandoning the Stirling approximation is similar to going backwards from integration to discrete summation.
In addition, numerical integration is performed via approximately finite summation, suggesting that numerical calculations can provide insight and methods for continuous integration of variation, which will be discussed further in the subsequent subsection.

\subsection{Hints from numerical analysis}\label{numer-intro}
The derivation of MPD involves both variational calculus and mathematical differences. In this article, we briefly introduce the essential concepts of these two subjects while also exploring potential applications for our study in statistical physics.

The calculus of variations, also referred to as variational calculus, is a field of mathematical analysis that utilizes variations, i.e. small changes in functions and functionals, to identify the minima and maxima of functionals. Functionals are commonly expressed as definite integrals involving functions and their derivatives, mapping from a set of functions to real numbers. Various techniques can be employed to determine the function that maximizes or minimizes the functional of a variational problem. Variational calculus is a significant branch of mathematics and plays a critical role in both classical and quantum physics. For instance, the application of variational calculus in classical mechanics entails the formulation of Hamilton's principle.

\begin{figure}[htbp]
\begin{center}
\includegraphics[width=8.5cm]{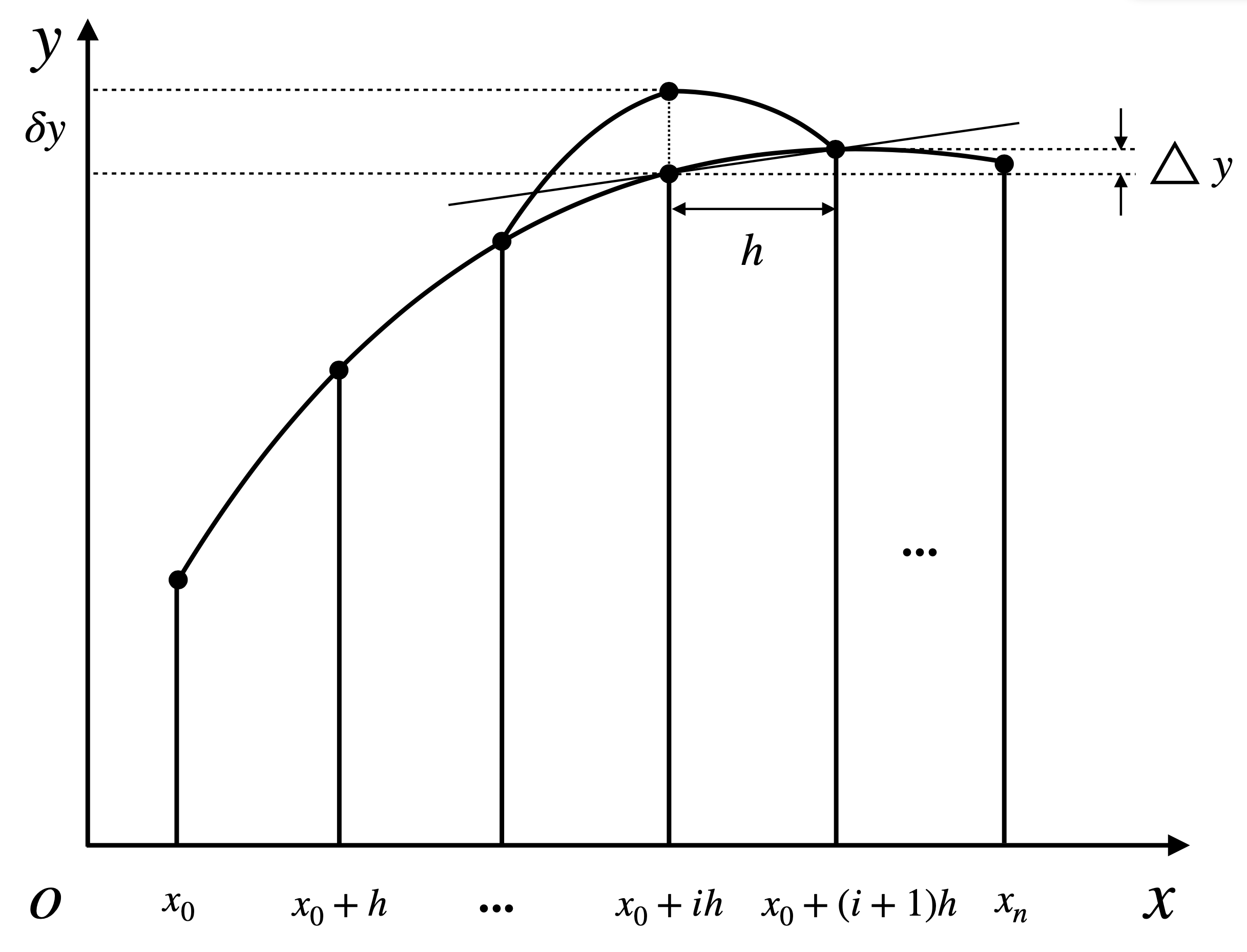} 
\includegraphics[width=8cm]{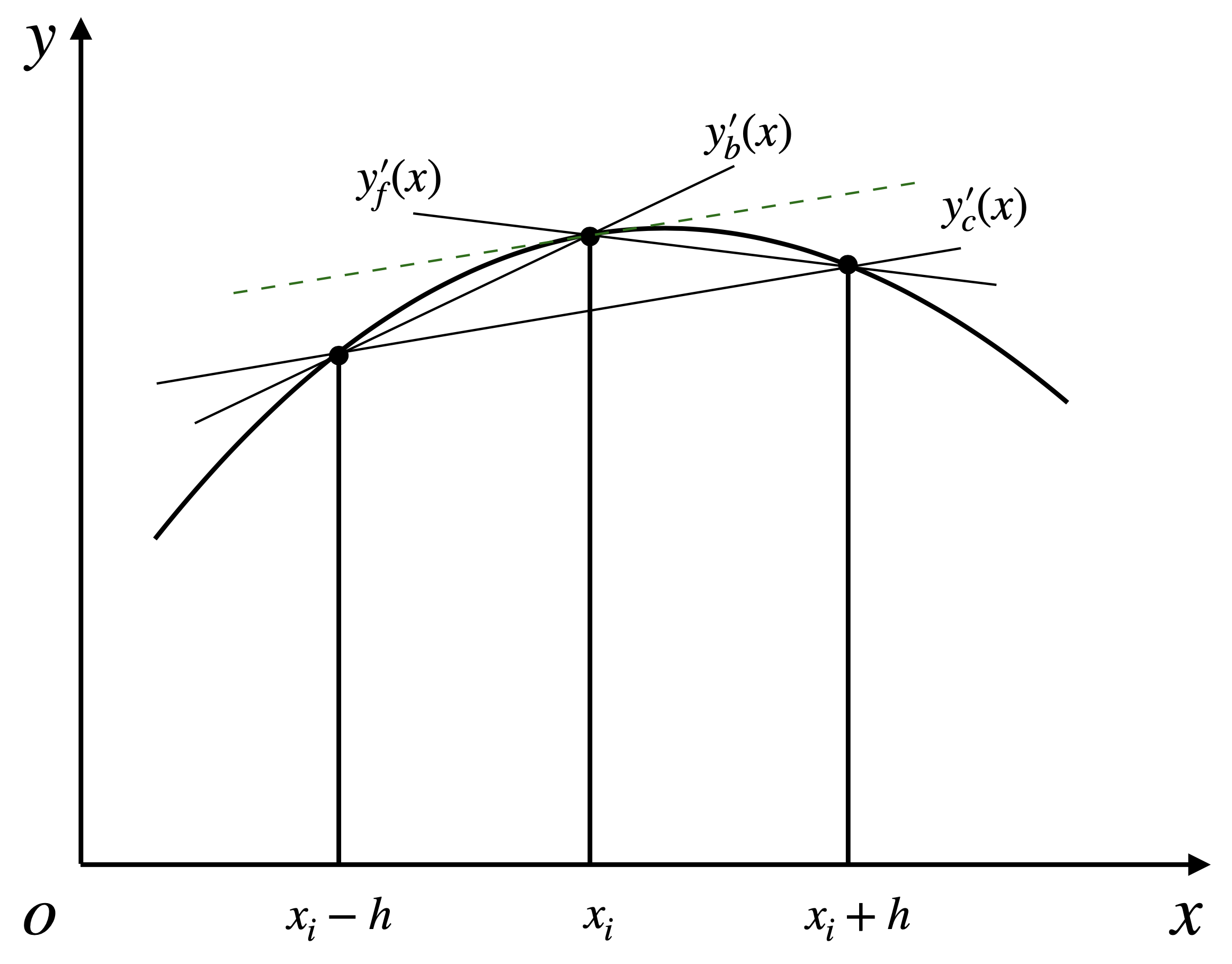} 
\end{center}
\caption{Left: Euler finite difference method for the
numerical variation. The difference between $\Delta y$ and
$y_{i+1}-y_i$ is shown in the figure.
Right: Geometric illustration of the forward,backward, 
and central finite difference formulas for approximating
$y^{\prime}(x_{i})$} \label{fig1:con}
\end{figure}

The direct method for solving variational problems is a crucial technique in variational calculus. It involves constructing a sequence of minimization and obtaining the solution to the problem through a limit process. The Euler finite difference method is a direct method \cite{Lao2021,hanc2004original,2003Calculus} that utilizes a piecewise linear function as the admissible function. With this type of function, the difference quotient and derivative are equivalent on the subinterval of the function segment. We investigate the functional
\begin{equation}
    J[y]=\int_{x_{0}}^{x_{n}}F(x,y,y')dx\, ,
\end{equation}
subject to the boundary
conditions $y(x_{0})=y_{0}$ and $y(x_{n})=y_{n}$. 
As shown in the left panel of Fig.~\ref{fig1:con}, 
the integral interval is
divided into $n$ subintervals of length $h$, with dividing
points at $x_{0}$, $x_{0}+h$, $\cdots$, $x_{0}+ih$,
$x_{0}+(i+1)h$, $\cdots$, $x_{0}+nh=x_{n}$. The
admissible function values at the dividing points are $y(x_{0})=y_{0}$,
$y_{1}$, $\cdots$, $y_{i}$, $y_{i+1}$, $\cdots$, $y_{n}=y(x_{n})$, where
$y_{1}$, $y_{2}$, $\cdots$, $y_{n-1}$ are undetermined. Thus, the
functional takes the following form,
\begin{eqnarray}
J[y]=\int_{x_{0}}^{x_{n}}F(x,y,y')dx\thickapprox\sum_{i=0}^{n-1}F(x_{i},y_{i},\frac{y_{i+1}-y_{i}}{h})h=\varphi(y_{1},y_{2},\cdots,y_{n-1})\,.
\end{eqnarray}
Finally, we solve the equation
\begin{equation}
 \frac{\partial\varphi}{\partial y_{i}}=0,~~
(i=1,2,\cdots,n-1)   
\end{equation}
 to obtain the minimum values of $y_{1}$,$y_{2}$,$\cdots$,$y_{n-1}$ for the function. By utilizing the Euler finite difference method, we transform the functional problem into a differential problem, which enables us to efficiently obtain the admissible function $y_{n}(x)$ with a concise linear score. The admissible curve, denoted by $y=(y_0(x),~y_1(x),\cdots,~ y_{n}(x))$, is a broken line that approximates the solution to the variational problem.
 
It is crucial to address a subtle point regarding Euler's method. The difference $\Delta y =y_{i+1}-y_i$ that yields the derivative $y'$ is not the finite variation $\delta y_i$ that is analyzed in the variation. As shown in the left panel of Fig.~\ref{fig1:con}, $\delta y_i$ refers to the variation of the single function $y_i$. It should be noted that $y_{i}$ can be continuous or discrete. The continuous case can be computed straightforwardly, while the discrete case can be obtained through difference methods that will be further discussed in the following context.

Next, it is necessary to re-examine the precision of the finite difference. Through a simple Taylor expansion in calculus, it can be demonstrated that the central finite difference formula exhibits superior accuracy compared to both the forward and backward finite difference formulas~\cite{nsde}.
An analytic function $y(x)$ may be written as a finite summation
\begin{eqnarray}
y(x+h)=y(x)+hy^{\prime}(x)+\frac{h^{2}}{2}y^{\prime\prime}(x)
+\cdots+\frac{h^{k}}{k!}y^{k}(\xi)\, .
\end{eqnarray}
If $y(x)$ is differentiable up to the $k$-th order, and $x<\xi$, the derivatives of a differential equation can be represented by finite difference formulas at grid points. This method results in a linear or nonlinear algebraic system. Various types of finite difference formulas are available, and their accuracy is directly proportional to the magnitude of $h$, which is typically small. It is worth noting that the forward finite difference and backward difference have the same accuracy order~\cite{nsde}. In this regard, we take the forward finite difference as an example.
\begin{eqnarray}
y^{\prime}_f (x_{i})\simeq\frac{y(x_{i}+h)-y(x_{i})}{h}
\sim y^{\prime}(x_i)\,. \label{eq:FD}
\end{eqnarray}
The forward finite difference method introduces an error which mandates the use of $h>0$. Geometrically, $y^\prime_f(x_{i})$ denotes the slope of the secant line that connects $(x_{i},y(x_{i}))$ and $(x_{i}+h,y(x_{i}+h))$, as depicted in the right panel of Fig.~\ref{fig1:con}. To evaluate the accuracy of $y^\prime_f(x_{i})$ in approximating $y^{\prime}(x_{i})$, we invoke the extended mean value theorem (Taylor series), provided that $y(x)$ has continuously differentiable second-order derivatives, as follows
\begin{eqnarray}
y(x_{i}+h)=y(x_{i})+y^{\prime}(x_{i})h
+\frac{1}{2}y^{\prime\prime}(\xi)h^{2}\,,
\end{eqnarray}
where $0<\xi<h$. Then the error can be estimated as 
\begin{eqnarray}
E_{f}(h)=\frac{y(x_{i}+h)-y(x_{i})}{h}
-y^{\prime}(x_{i})=\frac{1}{2}y^{\prime\prime}(\xi)h=O(h)\,.
\end{eqnarray}
As a result, the error, defined as the difference between the approximate value and the exact one, becomes proportional to $h$. The discretization equation Eq.~\eqref{eq:FD} is said to exhibit first-order accuracy.
The estimation of the central finite difference needs to retain 
more terms in the Taylor expansion 
\begin{eqnarray}
y(x+h)=y(x)+hy^{\prime}(x)+\frac{1}{2}y^{\prime\prime}(x)h^{2}+\frac{1}{6}
y^{\prime\prime\prime}(\xi )h^{3}\,,
\\
y(x-h)=y(x)-hy^{\prime}(x)+\frac{1}{2}y^{\prime\prime}(x)h^{2}-\frac{1}{6}
y^{\prime\prime\prime}(\xi)h^{3}\,,
\end{eqnarray}
which lead to 
\begin{eqnarray}
E_{c}(h)=\frac{y(x_{i}+h)-y(x_{i}-h)}{2h}-y^{\prime}(x_{i})=\frac{1}{6}y^{\prime\prime\prime}(\xi )h^{2}=O(h^{2})\,.
\end{eqnarray}
Thus the central finite
difference formula is the second-order accuracy. 

As discussed earlier, the logic behind numerical variation and error analysis involves returning from continuous calculus to discrete summation. Interestingly, discrete summations are also the initial evaluation of MPD in statistical physics. Continuous distribution functions are merely a limit with infinitely large particle numbers. Therefore, it is natural to apply numerical variation and error analysis to deal with finite number statistical systems. The Euler finite difference method can be used for variation to give the maximum distributions. Furthermore, central difference should be adopted for difference derivation as it provides a more accurate estimation. We will provide further details in the next section.

\section{The  most probable distribution 
with finite particle numbers}\label{sec3}
\subsection{Variation of finite number system}

Inspired by numerical analysis, we can provide better treatment for the most probable distribution in statistical physics. For example, let us consider Maxwell-Boltzmann distribution\cite{greiner2012thermodynamics}.
 The logarithm of the distinct microstates in a set
$\{n_i\}$ is
\begin{eqnarray}
\ln\Omega_{\rm M. B.}=\ln\biggl(\frac{N!}{\prod_{i}n_{i}!}\prod_{i}g_{i}^{n_{i}}\biggl)=\ln N!+\sum_{i}n_{i}\ln g_{i}-\sum_{i}\ln n_{i}!\,.
\end{eqnarray}
The equation above represents a physical system with a 
finite number of particles. Only when $n_i \gg 1$ for every energy level can the logarithm be treated as a continuous integration using Stirling's approximation. 
To better understand the difference between ``discrete" 
and ``continuous", 
we can assume that $n_i$, $g_i$, and $\varepsilon_i$ are continuous functions of energy level $i$. This allows us to transform the summation of subscript $i$ 
into an integration over a continuous variable $x$.
The logarithm of the number of the microstates
will be 
\begin{eqnarray}
\ln\Omega_{\rm M. B.}
=C+\int\biggl[\tilde{n}(x)\ln\tilde{g}(x)
-\ln\Gamma(\tilde{n}(x)+1)\biggl]{\rm d}x\,,
\end{eqnarray}
where $C=\ln N!$. 
Note that the symbol ``$~\tilde{\ }~$" is added in 
$\ensuremath{\tilde{n}(x)}$, $\ensuremath{\tilde{g}(x)}$, and
$\ensuremath{\tilde{\varepsilon}(x)}$ for the distinguishment with 
the discrete variables. 
The total number and total energy constraints of the system will be 
\begin{eqnarray}
\int\tilde{n}(x){\rm d}x=N\,, \mbox{\ and \ } \int\tilde{n}(x)\tilde{\varepsilon}(x){\rm d}x=E\,.
\end{eqnarray}
Now we define functional $J$ with  $\alpha$ and $\beta$ 
as Lagrangian multipliers
\begin{eqnarray}
J(\tilde{n})=\int\left[\tilde{n}(x)\ln\tilde{g}(x)-\ln\Gamma(\tilde{n}(x)+1)-\alpha\left(\tilde{n}(x)-N\right)-\beta\left(\tilde{n}(x)\tilde{\varepsilon}(x)-E\right)\right]{\rm d}x\,.
\end{eqnarray}
This is a continuous functional of the distribution $\tilde{n}(x)$. The variation of $\delta J=0$ yields the most probable distribution,
which is the Maxwell-Boltzmann distribution in this case.

\begin{figure}[htbp]
\centering
    \includegraphics[width=0.45\linewidth]{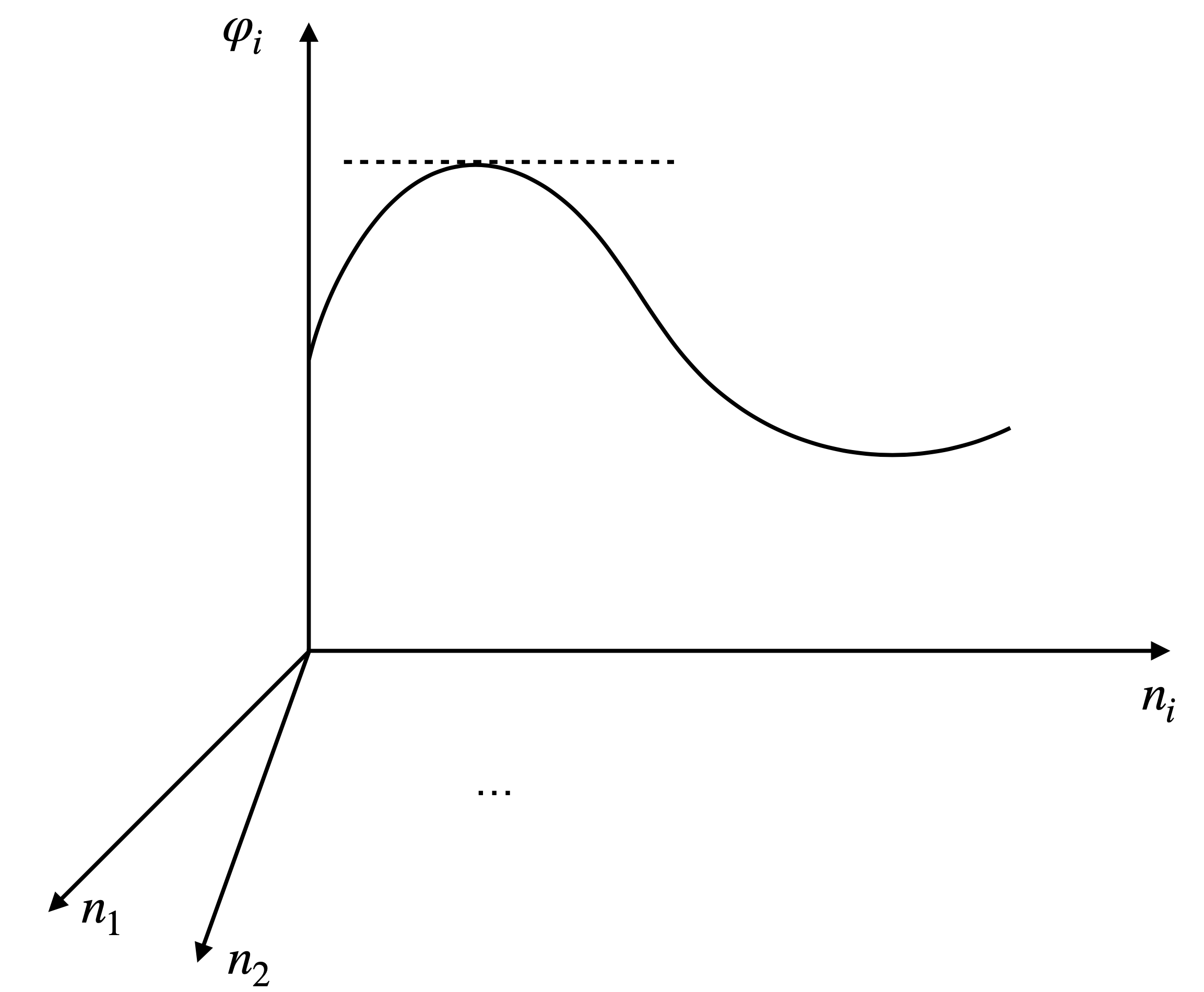}
    \includegraphics[width=0.45\linewidth]{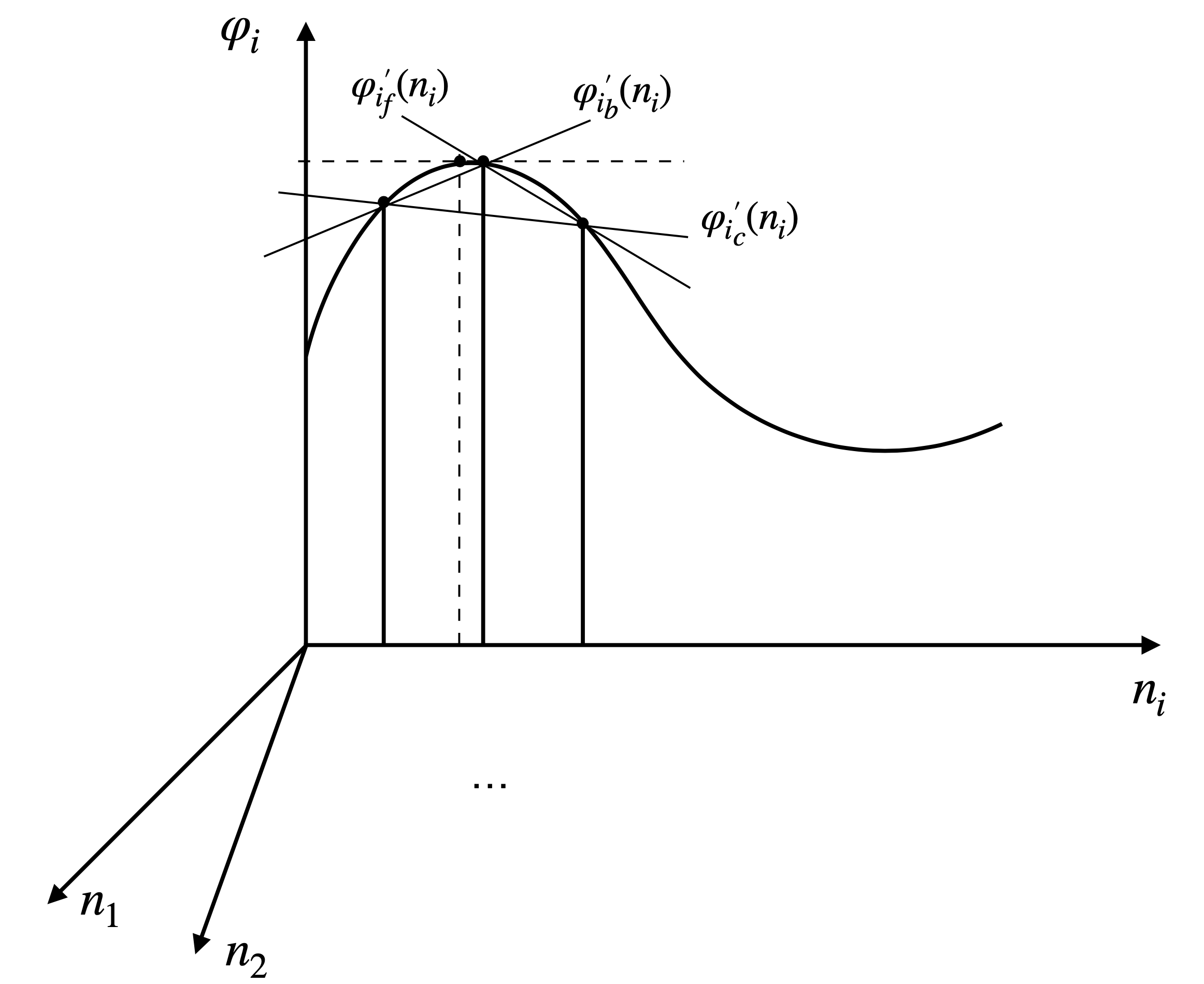}
\caption{
Left: the maximum of a continuous functional; Right:
the maximum value of a discrete functional.}\label{fig2:d}
\end{figure}

Next the Euler finite difference method is used for the finite
particle number. All the function such as $\ensuremath{\tilde{n}(x)}$
return to their normal form. The functional will be
\begin{eqnarray}
J(n_{i})&=&\sum_{i}n_{i}\ln
g_{i}-\sum_{i}\ln\Gamma(n_{i}+1)+\alpha\left(\sum_{i}n_{i}-N\right)
-\beta\left(\sum_{i}n_{i}\varepsilon_{i}-E\right)\nonumber\\
&=&\Phi\left[n_{1},n_{2},\cdots,n_{i_{max}};g_{i},
\varepsilon_{i},N,E\right]\,.
\end{eqnarray}
Here $i_{max}$ denote the largest energy level of the system, which
could be any sufficiently large number.
At this point, we should note that 
$n_{1},n_{2},\cdots,n_{i_{max}}$
form a mutually function
space as depicted in Fig.~\ref{fig2:d}.
The extreme value of the functional $J(n_{i})$
could be derived  by the requirement that the variation of $\Phi$
on every $n_i$ equal zero
\begin{equation}
    \ensuremath{\frac{\dot \delta \Phi}{\dot \delta  n_{i}}
    \dot =0}\, .\label{varPhi}
\end{equation}
Here the symbol ``$\dot \delta $" and ``$\dot =$"
are used to represent the discrete variation and equality. 
$\dot \delta n_i$ 
means that it has the minimum interval, which is
\begin{equation}
    \dot \delta n_i  =1\,.
\end{equation}
Naturally, $\dot{\delta}\Phi$ is also finite, in contrast to continuous variation which is infinitesimal. Another subtle point to note is the equivalence to zero. For a continuous functional, the variation is exactly equal to zero, as demonstrated in the left panel of Fig.\ref{fig2:d}. However, as shown in the right panel of Fig.\ref{fig2:d}, the equality can only be maintained within a certain degree of uncertainty for discrete variation. The newly defined symbol $\dot{=}$ reminds us to identify the point closest to ${\dot\delta\Phi}/{\dot\delta n_i}=0$. Furthermore, the degree of uncertainty is dependent on the specific differences involved. As discussed earlier, the central difference has a higher order of accuracy as compared to the forward and backward differences. Therefore, the central difference method is adopted for the subsequent studies.

To give a uniform description of the variation, 
the total number and total energy are divided into 
\begin{equation}
    N_{min}=N/i_{max},~~~~E_{min}= E/i_{max}\,.
\end{equation}
Then the functional can be written as
\begin{eqnarray}
\Phi = \sum_{i} \varphi_i
= \sum_{i}\left[n_{i}\ln g_{i}-\ln\Gamma(n_{i}+1)
-\alpha\left(n_{i}-N_{min}\right)-\beta\left(n_{i}
\varepsilon_{i}-E_{min}\right)\right]\,.
\end{eqnarray}
The extreme variation  of Eq.~\eqref{varPhi} gives equations
\begin{eqnarray}
\dot \delta \varphi_i \dot = 0\,.\label{varpi}
\end{eqnarray}
The central difference is
\begin{eqnarray}
\triangle_c\varphi_{i}&=&
\left[(n_{i}+1)\ln g_{i}-\ln\Gamma(n_{i}+2)-\alpha\biggl(n_{i}+1-N_{min}\biggl)-\beta\biggl((n_{i}+1)\varepsilon_{i}-E_{min}\biggl)\right] 
\nonumber\\
&& -\biggl[(n_{i}-1)\ln g_{i}-\ln\Gamma(n_{i})-
\alpha\biggl(n_{i}-1-N_{min}\biggl)-\beta\biggl((n_{i}-1)\varepsilon_{i}-E_{min}\biggl)\biggl] \,.
\end{eqnarray}
Then extreme requirement of Eq.~\eqref{varpi} gives
\begin{eqnarray}
\frac{n_{i}+1}{g_{i}}\frac{n_{i}}{g_{i}}\text{\ensuremath{\dot
=}}e^{-2(\alpha+\beta\varepsilon_{i})}\,.
\label{eq:Boltzmann}
\end{eqnarray}
It is obvious that the distribution will recover the classical
Maxwell-Boltzmann distribution in case of $n_i\gg 1$
\begin{equation}
    n_i = g_i e^{-\alpha-\beta\varepsilon_{i}}\,.
\end{equation}
However, if the number of particles is finite, we must acknowledge that the distribution exists in an implicit form. If the total number scale is significantly less than Avogadro's constant, the deviation from the ordinary Maxwell-Boltzmann distribution may be significant enough to detect. Compared with the asynchronous difference method in Ref.~\cite{Liu_2022}, we observe that the central difference discards the division of $y$ into $y_1$ and $y_2$, which had varying differences. The central difference further enhances accuracy.
Additionally, we note the significance of uncertainties being equated to zero. The coefficients, Lagrangian multipliers $\alpha$ and $\beta$, are consistent across all energy levels, determined by temperature and chemical potential. Imposing equality between these coefficients and an integer number $n_i$ is unphysical, and thus our work defines an equivalence with some degree of uncertainty.
Moreover, as $n_i$ has a lower boundary ($n_i\ge 0$), it is noteworthy that the extreme of variation may be at said boundary $n_i=0$,
\begin{equation}
\dot \delta \varphi_i < 0\,,
\end{equation}
although studying such an eventuality surpasses the scope of our work.

In the same way, we can get Bose-Einstein distribution
\begin{eqnarray}
\frac{(n_{i}+1)n_{i}}{(n_{i}+g_{i})(n_{i}+g_{i}-1)}
{\dot = }e^{-2(\alpha+\beta\varepsilon_{i})}\,, \label{eq:bose}
\end{eqnarray}
and Fermi-Dirac distribution
\begin{eqnarray}
\frac{(n_{i}+1)n_{i}}{(g_{i}-n_{i})(g_{i}-n_{i}+1)}
{\dot =}e^{-2(\alpha+\beta\varepsilon_{i})}\, . \label{eq:fermi}
\end{eqnarray}
Similar ordinary distributions can be achieved when $n_i$ is significantly large. Please note that the ordinary distributions are referred to as canonical distributions while the newly derived distributions through finite difference are denoted as finite distributions in the subsequent context.

All our results of the MPD from canonical and finite
distributions are listed in the following tabular for the comparison.
\begin{center}
\begin{tabular}{c|c|c}
\hline\hline
 & canonical distributions  & finite distributions   \\ \hline
    Maxwell-Boltzmann distribution &
    $n_{i}=g_{i}e^{-(\alpha+\beta\varepsilon_{i})}$ &
$\frac{n_{i}+1}{g_{i}}\frac{n_{i}}{g_{i}}\text{\ensuremath{\dot
=}}e^{-2(\alpha+\beta\varepsilon_{i})}$ \\ \hline
    Bose-Einstein distribution &  $n_{i}=\frac{g_{i}}{e^{\alpha+\beta\varepsilon_{i}}-1}$
& $\frac{(n_{i}+1)n_{i}}{(n_{i}+g_{i})(n_{i}+g_{i}-1)}
{\dot = }e^{-2(\alpha+\beta\varepsilon_{i})}$   \\ \hline
    Fermi-Dirac distribution &  $n_{i}=\frac{g_{i}}{e^{\alpha+\beta\varepsilon_{i}}+1}$
& $\frac{(n_{i}+1)n_{i}}{(g_{i}-n_{i})(g_{i}-n_{i}+1)}
{\dot =}e^{-2(\alpha+\beta\varepsilon_{i})}$\\
    \hline\hline
\end{tabular}
\end{center}
Of course, the differences between canonical distributions and finite distributions are negligible when the particle numbers are infinitely large. However, as discussed in the introduction, the number scales of many physical systems are much less than Avogadro's constant, which means that observable phenomena may arise due to these differences. This is especially true for Bose-Einstein condensation found in cold atom systems. Finite distributions may be applicable, and the details will be studied in the following subsection.

\subsection{Application in Bose-Einstein condensation.}

Modifications of canonical distributions are derived using central difference. In this subsection, we use these distributions to study Bose-Einstein condensation. First, the sketch picture with continuous distribution is reviewed~\cite{huang2008statistical}. Consider a system consisting of $N$ identical and nearly independent bosons with temperature $T$ and volume $V$. In the case of an infinite number of particles, the distribution is canonical with chemical potential $\mu $ and temperature $T$
\begin{eqnarray}
\alpha=-\frac{\mu}{kT},~~\beta=\frac{1}{kT}\,,
\end{eqnarray}
 namely 
\begin{eqnarray}
n_i=\frac{g_i}{e^{\frac{\varepsilon_i-\mu}{kT}}-1}\,.\label{fibo}
\end{eqnarray}
The key point of the distribution is the $-1$ in the denominator, which leads to condensation in momentum space. 

Suppose the lowest energy level is zero $\varepsilon_0=0$, the condensation occurs at the critical temperature $T_c$ where the chemical potential $\mu$ vanishes, with the critical number distribution given by
\begin{eqnarray}
n_c=\frac{N}{V}= \frac{1}{V}\sum_i\frac{g_i}
{e^{\frac{\varepsilon_i}{kT_c}}-1}\,. \label{crden}
\end{eqnarray}
Periodic boundary conditions for the de Broglie wave at the wall are used to estimate the energy levels. If a free particle is located in a cubic container with edge length $L$, the possible values of energy for the three-dimensional free particle are given by
\begin{eqnarray}
\varepsilon=\frac{1}{2m}(p_{x}^{2}+p_{y}^{2}+p_{z}^{2})=\frac{2\pi^{2}
\hbar^{2}}{m}\frac{n_{x}^{2}+n_{y}^{2}+n_{z}^{2}}{L^{2}}\, ,\quad\quad
n_{x},n_y, n_z=0,\pm1,\pm2,\cdots\,\label{epsn}
\end{eqnarray}
$n_{x}$, $n_{y}$, and $n_{z}$ are the quantum numbers used to characterize the motion state of a three-dimensional free particle. It should be noted that there are two propagation directions for waves, therefore $n_{x}$, $n_{y}$, and $n_{z}$ can be negative integers. The energy level is dependent solely upon $n_{x}^{2}+n_{y}^{2}+n_{z}^{2}$, and as a result, degeneracy is always going to be greater than one. The number of quantum states for a free particle within the momentum range from $p_{x}$ to $p_{x}+dp_{x}$, $p_{y}$ to $p_{y}+dp_{y}$, and $p_{z}$ to $p_{z}+dp_{z}$, inside a volume $V=L^{3}$ can be described as follows
\begin{eqnarray}
dn_{x}dn_{y}dn_{z}=\left(\frac{L}{2\pi\hbar}\right)^{3}dp_{x}dp_{y}dp_{z}=
\frac{V}{h^{3}}dp_{x}dp_{y}dp_{z}\,\label{dndp}\,.
\end{eqnarray}
Use the relation $\varepsilon = p^2 /2m$, we can obtain 
the number of possible states of a free particle within 
the energy range of $\varepsilon$ to $\varepsilon+d\varepsilon$ 
in the volume $V$
\begin{eqnarray}
D(\varepsilon)d\varepsilon=\frac{2\pi V}{h^{3}}
(2m)^{3/2}\varepsilon^{1/2}d\varepsilon\,.\label{condis}
\end{eqnarray}
In the case of an infinitely large number of particles, the energy level spacing is much smaller than $kT$, so we can replace the summation in Eq.~\eqref{crden} with an integral, treating the energy levels as continuous. The number of possible states of a free particle within the energy range of $\varepsilon$ to $\varepsilon+d\varepsilon$, or the degeneracy in that energy range, is given by Eq.~\eqref{condis}.
Thus the expression becomes
\begin{eqnarray}
n_c=\frac{2\pi}{h^{3}}(2m)^{3/2}\int_{0}^{\infty}\frac{\varepsilon^{1/2}d\varepsilon}{e^{\frac{\varepsilon}{kT_c}}-1}
= \frac{2\pi}{h^{3}}(2mkT_{c})^{3/2}\int_{0}^{\infty}\frac{x^{1/2}dx}{e^{x}-1}=\left(\frac{2\pi mkT_{c}}{h^2}\right)^{3/2}\times2.612
\, \label{ncTc}\,. 
\end{eqnarray}
Therefore, the relation between  critical number density $n_c$  and 
the critical temperature $T_{c}$ is
\begin{eqnarray}
T_{c}=\left(\frac{n_c}{2.612}\right)^{2/3}
\frac{2\pi \hbar^{2}}{mk}\, .\label{Tcnc}
\end{eqnarray}
As the temperature continues to decrease, 
a macroscopic number of particles will condense
at the lowest energy
level $\varepsilon=0$ of which the density is 
\begin{eqnarray}
n_{0}(T)=n_c\left[1-\left(\frac{T}{T_{c}}
\right)^{3/2}\right]\, .\label{n0nc}
\end{eqnarray}
The collection of particles condensing in $\varepsilon_{0}$ 
is known as a Bose-Einstein condensate.
This indicates that, below critical temperature $T_{c}$, $n_{0}$ and $n$
have the same order of magnitude.
Rewriting the Eq.~\eqref{Tcnc} as 
\begin{eqnarray}
n\left(\frac{h}{\sqrt{2\pi mkT_{c}}}\right)^{3}
=n\lambda^{3}=2.612\,,\label{denlam}
\end{eqnarray}
where $\lambda$ is the thermal wavelength of the atoms, which is larger
than the average distance between atoms. The Eq.~\eqref{denlam} gives
the critical condition for the appearance of Bose-Einstein condensation
in an ideal gas, and the condition for the presence of a condensate 
is $n\lambda^{3}\geq2.612$.

The exploration of condensation with finite number distributions should begin with counting the energy levels. The smallest energy interval should be determined by the edge length $L$ or, more specifically, the volume $V$. From the possible energy level given in Eq.~\eqref{epsn}, the energy interval can be written as
\begin{eqnarray}
\Delta\varepsilon =\frac{2\pi^2\hbar^2 }{mV^{2/3}}\,.
\end{eqnarray}
We can see that the interval is proportional  to the $V^{-2/3}$.
When we change the integral of Eq.~\eqref{ncTc} to the summation
of discrete energy level, not like the Eq.~\eqref{dndp}, the volume $V$ cannot be directly extracted out of $dn_x dn_y dn_z$.
However, the critical density can be simply derived by the summation of 
finite distribution. When  the distribution is canonical
\begin{eqnarray}
n_c^{\rm C} = \left(\sum_{n_x,n_y,n_z}\frac{1}{
e^{\frac{\Delta \varepsilon}{kT_c }(n_x^2 +n_y^2
+n_z^2 )}-1}\right)\!\Biggl/V\,,\label{nccan}
\end{eqnarray}
which is distinguished by a superscript $\rm C$.
When the distribution is finite
\begin{eqnarray}
n_c^{\rm F} = \left(\sum_{n_x,n_y,n_z}F(g_i,n_x,n_y,n_z)\right)
\!\Biggl/V\,,\label{ncfin}
\end{eqnarray}
which is distinguished by a superscript $\rm F$.
$F(g_i,n_x,n_y,n_z)$ is the explicit form of the finite distribution
which can be derived by solve the quadratic equation Eq.~\eqref{eq:bose}
\begin{eqnarray}
F(g_i,n_x,n_y,n_z)
=\frac{\sqrt{4g_{i}^{2}e^{2(\alpha+\beta\varepsilon_{i})}-8g_{i}e^{2(\alpha+\beta\varepsilon_{i})}+e^{4(\alpha+\beta\varepsilon_{i})}+2e^{2(\alpha+\beta\varepsilon_{i})}+1}+2g_{i}-e^{2(\alpha+\beta\varepsilon_{i})}-1}{2\left(e^{2(\alpha+\beta\varepsilon_{i})}-1\right)}\,,
\end{eqnarray}
with $\alpha =0$ and $\beta\varepsilon_i = 
({\Delta \varepsilon}/{kT_c })(n_x^2 +n_y^2 +n_z^2 )$.
Note that here we change  ``$\dot =$" as ``$=$" for simplicity.
One can easily check that in case of ${\Delta \varepsilon}/{kT_c }\ll 1$,
the critical density will be equal to the density got by integration of
the continuous distribution Eq.~\eqref{ncTc}. 
When ${\Delta \varepsilon}/{kT_c }$ are comparable to $1$ or 
sufficiently large,
the density will be changed by deviation between discrete summation 
and the continuous integration. 

\begin{figure}[htbp]
\begin{center}
\scalebox{0.6}{\epsfig{file=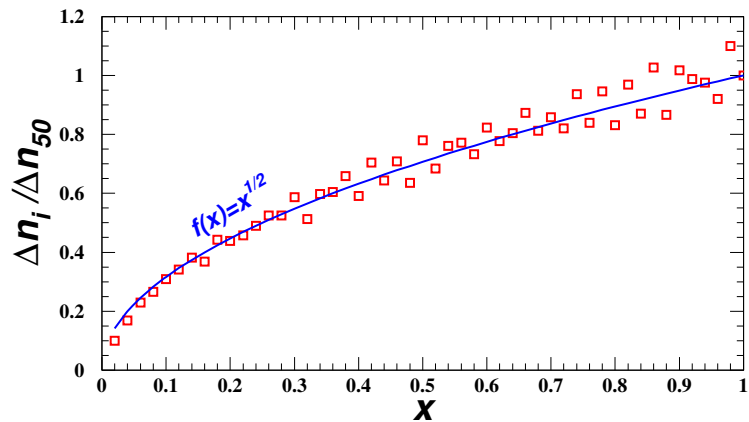} }
\end{center}
\caption{ Illustration of the deviation of the number density 
between discrete summation and the continuous function $\sqrt{x}$.}
\label{figdn}
\end{figure}

\begin{figure}[htbp]
\begin{center}
\scalebox{0.6}{\epsfig{file=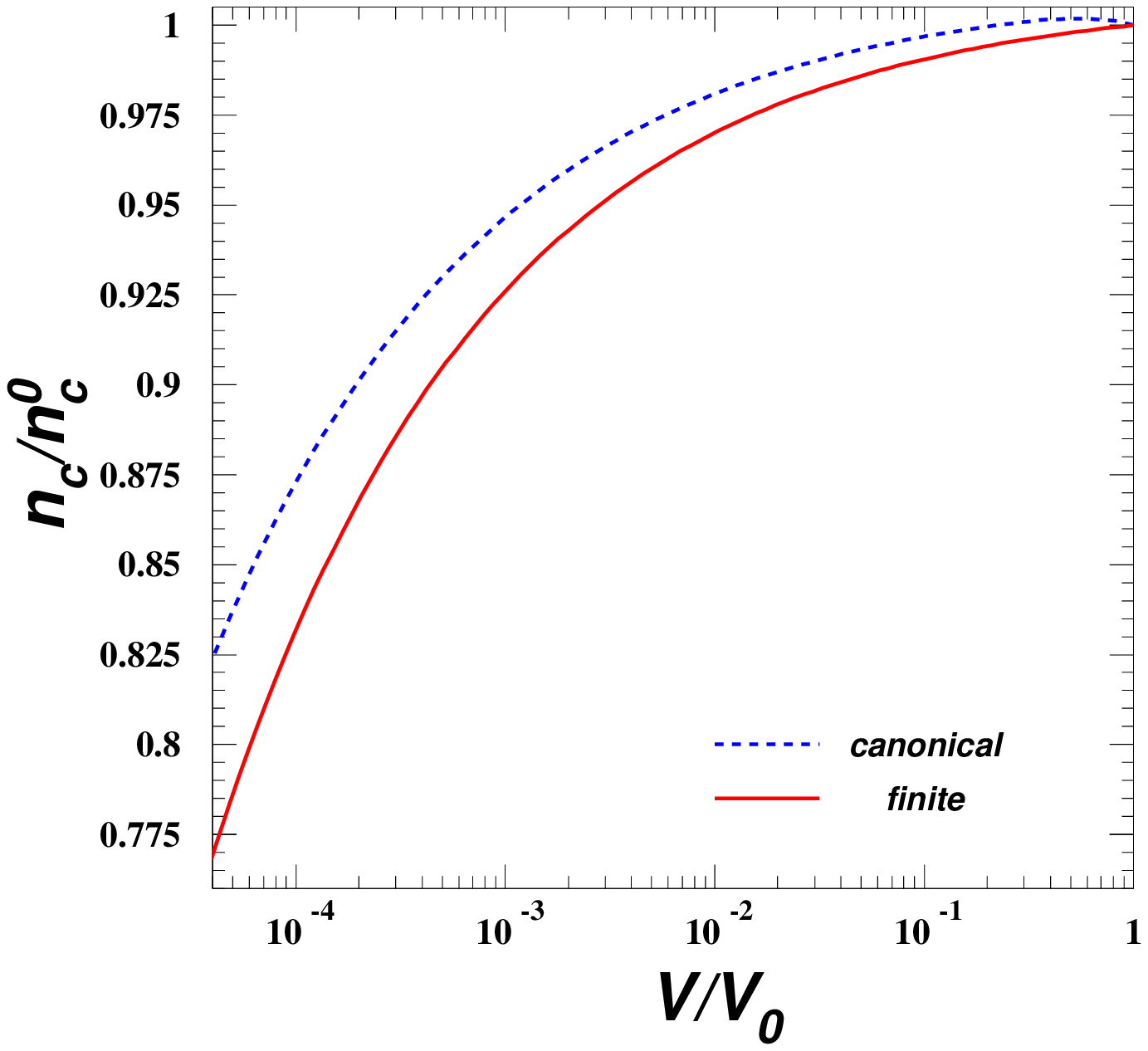} }
\raisebox{4mm}{\scalebox{0.59}{\epsfig{file=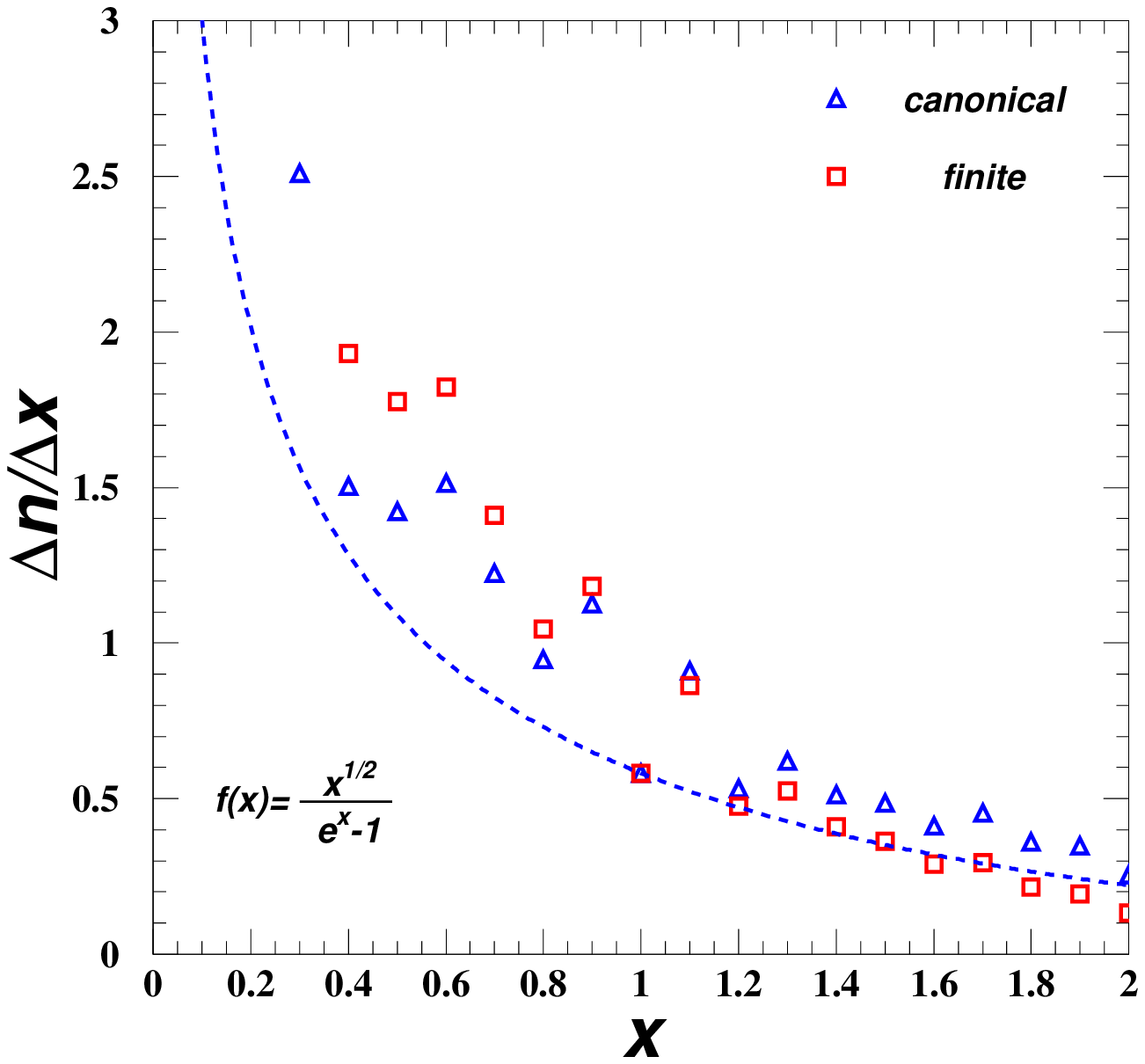} }}
\end{center}
\caption{Left: the change of the critical density along with
the volume, under the assumption of a constant critical temperature.
Right: The normalized canonical and finite distributions
in the the energy space.} \label{figncdis}
\end{figure}

In fact, from the number density given in Eq.~\eqref{condis}, the density obeys a square root $\sqrt{x}$ 
relation with energy. To show the deviation from continuous integration, we recorded the number of states in intervals of 20 in the number space up to 1000; for example, $\Delta n_{10}$ records the states between $180-200$. We then normalized all the $\Delta n_i$ with
$\Delta n_{50}$ and compared it with the function $\sqrt{x}$. The results are shown in Fig.~\ref{figdn}. From the figure, we can see that although the tendency of the discrete density is $\sqrt{x}$,  there is an obvious deviation.
Thus, we assume that the critical temperature remains constant and set
${\Delta \varepsilon}/{kT_c }=10^{-4}$ as a reference critical density $n_c^0$ with volume $V_0$. 
We then vary ${\Delta \varepsilon}/{kT_c }$
from $10^{-4}$ to $10^{-1}$, 
which also means that the volume of the system $V/V_0$
(normalized by  $V_0$)  varies  from $1$ to $10^{-4.5}$.
The corresponding results of $n_c/n_c^0$ are shown in the left panel of 
Fig.~\ref{figncdis}. Note that, we chose $g_i=3$ in the numerical
calculation of finite distributions. One can check that 
finite distribution $F(g_i,n_x,n_y,n_z)=0$ when $g_i =1$.
$F(g_i,n_x,n_y,n_z)$ will exactly match the canonical distribution when $g_i=2$. Thus we chose a universal $g_i=3$ 
for demonstration purposes. Additionally, note that the summation of $n_x,n_y,n_z$
ranges from a large negative integer $-M$ to $M$, so the degeneracy resulting from changes in sign is already taken into account in the summation.
From the figure, we can see that as the volume decreases, the critical densities also decrease to significant values. The decrease of the finite distribution is larger than that of the canonical distribution. For example, at the critical density point $V/V_0=10^{-4.5}$, the canonical critical density decreases to 0.825 of the density derived from continuous distribution, while the finite critical density decreases to 0.76. The difference between the two may be detected in future measurements on Bose-Einstein condensation.
The distribution of states in the energy space is shown in the right panel of Fig.~\ref{figncdis}, in which 
$\Delta \varepsilon/kT_C= 10^{-2}$. 
The canonical and finite distributions are compared with the continuous function  $f(x)=\sqrt{x}/(e^x-1)$ (see Eq.~\eqref{ncTc}).
We can see that there are significant deviations from continuous distributions, which is the reason for the decrease in critical density. This decrease implies that the critical condition Eq.~\eqref{denlam} equivalent to 2.612 can be relaxed to a smaller value. For example, if we take $\Delta \varepsilon/kT_c =10^{-1}$
, the decrease in critical density means that the condensation occurs at the point 
$n\lambda^3=0.612\times 0.76=1.99$. This is an important feature that may serve as verification of the finite distribution.
Next, we will analyze the phenomenology of Bose-Einstein condensation experiments in different distributions.

\begin{figure}[htbp]
\begin{center}
\scalebox{0.65}{\epsfig{file=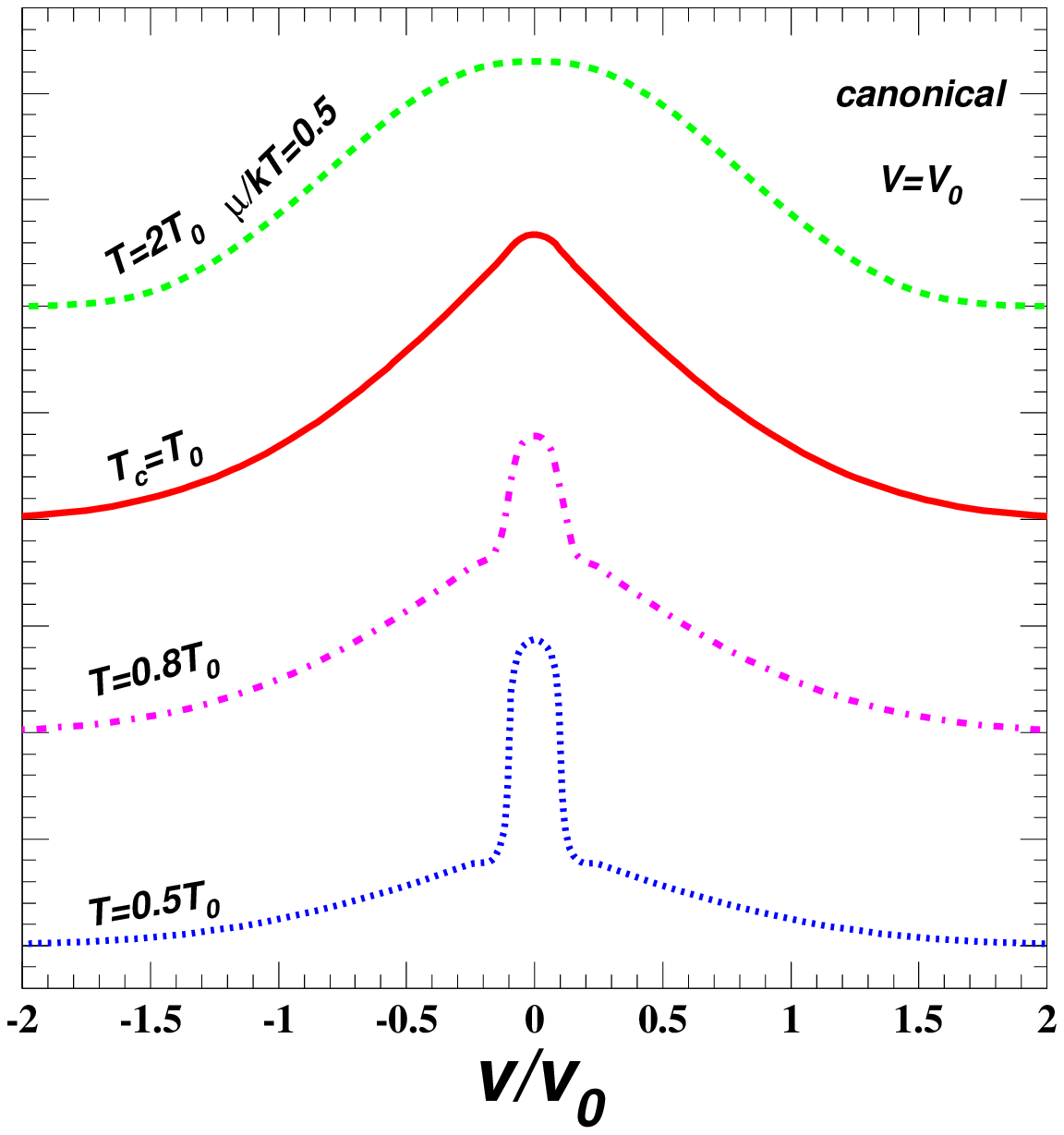} }
\scalebox{0.65}{\epsfig{file=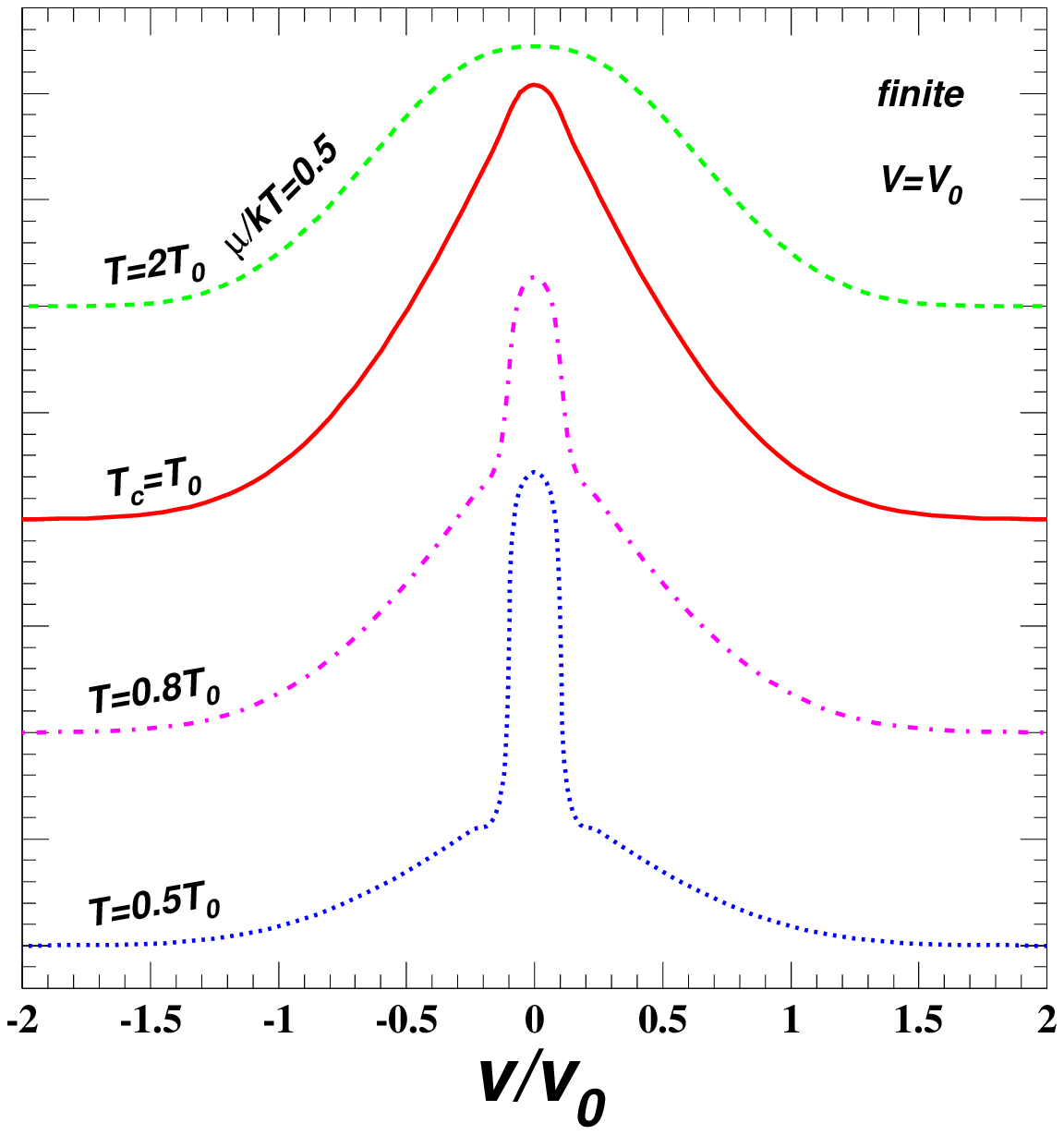} }
\scalebox{0.65}{\epsfig{file=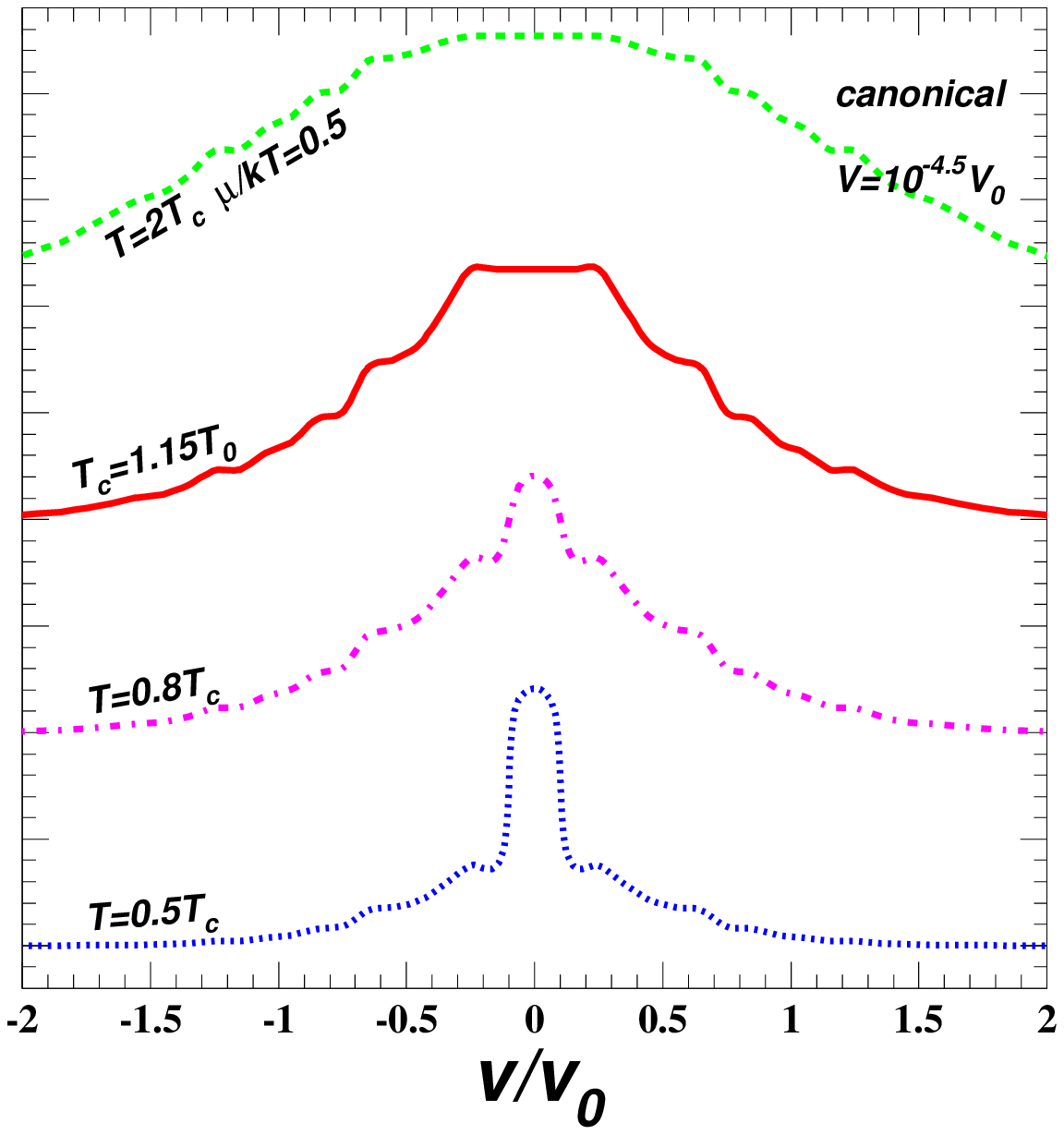} }
\scalebox{0.65}{\epsfig{file=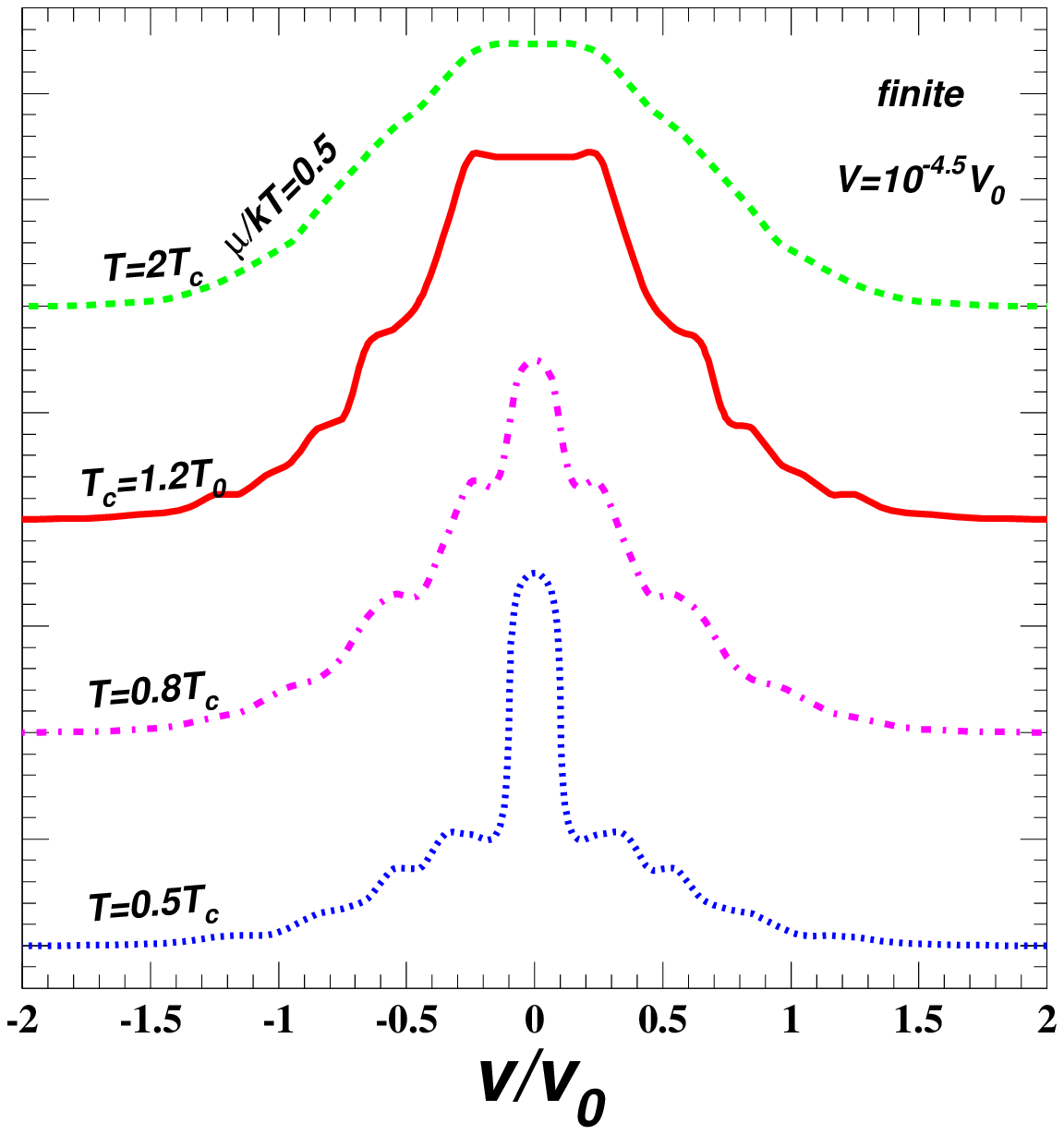} }
\end{center}
\caption{The velocity distributions of Bose-Einstein condensation.
Upper left: the condensation of canonical distributions
at different temperature for the matter with a volume $V_0$;
Upper right: the condensation of finite distributions
at different temperature for the matter with a volume $V_0$;
Lower left: the condensation of canonical distributions
at different temperature for the matter with a volume 
$10^{-4.5}V_0$;
Lower right: the condensation of finite distributions
at different temperature for the matter with a volume 
$10^{-4.5}V_0$.}\label{figvedis}
\end{figure}

Ever since the 1980s, breakthroughs have been made in laser cooling,
magneto-optical traps, and evaporative cooling technology. 
Bose-Einstein condensation of $^{87}\rm Rb$~\cite{anderson1995observation},
$^{23}\rm Na$\cite{davis1995bose}, and $^{7}\rm Li$~\cite{bradley1995evidence} vapors
was achieved in 1995. 
Take $^{23}\rm Na$ condensation as an example,
the rubidium atoms in the background
vapor were optically pre-cooled and captured, then loaded into a magnetic
trap and further cooled by evaporation. The evaporation cooling is
achieved by releasing higher energy atoms from the trap through a radio
frequency (rf) magnetic field, so that the remaining atoms reach a lower
temperature. By setting the frequency of the rf field, the atoms can
be selectively driven to untrapped spin states. For optimum cooling, the rf frequency is slowly lowered, which increases the center
density and collision rate, leading to temperature reduction. 
Bose-Einstein condensation can be observed by measuring the velocity distribution of the evaporative cloud, as detailed in Ref.~\cite{anderson1995observation}. We used canonical and finite distributions to produce velocity distributions at different particle number scales, simulating the velocity distributions in Ref.~\cite{anderson1995observation}. Other numerical calculations were conducted as follows.
First, a volume $V_0$ is chosen, and we assume it gives $\Delta\varepsilon/kT_c = 10^{-4}$ to simulate the continuous distribution. The critical temperature is chosen as $T_c = T_0$. Then, the velocity distributions are numerically counted in both canonical and finite distributions with temperatures chosen as  $2T_0,~T_0,~0.8T_0,~0.5T_0$.
 The numerical results are shown in the upper two panels of Fig.~\ref{figvedis}. Note that we chose the chemical potential as $\mu/kT_c=0.5$ for the distributions at temperatures higher than the critical temperature, and the reference velocity $v_0$ 
is defined as $\sqrt{2mkT}$. 

From these panels, we can see that at temperatures higher than the critical temperature, the distributions are similar to each other. After passing through the critical temperature, a narrow peak arises, showing the condensed particles in the zeroth energy level. The smooth broad distribution shows particles in other higher energy levels. Comparing the two distributions, we find that the finite distribution gives a sharper distribution in the velocity space. Additionally, the number of particles in the condensate state at the zeroth energy level is obviously larger in the finite distribution than in the canonical distribution.

Next, we changed the volume to $10^{-4.5}V_0$, which means that
$\Delta\varepsilon/kT=10^{-1}$. We then counted the velocity distributions again using both canonical and finite distributions, with the results shown in the lower panels of Fig.~\ref{figvedis}. As discussed above, the critical condition $n\lambda^3 = 2.612$ is broken down. Since 
the $\lambda $ is determined by temperature, if the trapped matter maintains a constant number density, the critical temperature is enhanced. Using the $n^c/n_c^0$ ratio calculated in Fig.~\ref{figncdis}, we can determine that the critical temperature for canonical distribution is $1.15T_0$,
and the critical temperature for finite distribution is $1.2T_0$.
From these panels, we can see that the distributions are also similar when the temperature is higher than the critical temperature.
However, when the temperature passes through the critical temperature, the distributions become step functions. This is due to a larger interval $\Delta\varepsilon/kT$
 in the denominator of the distributions Eq.~\eqref{nccan} and Eq.~\eqref{ncfin}. Only energy levels that satisfy the special number constraints  $n_x^2 +n_y^2+n_z^2$ can exist. Because the energy interval is sufficiently large,
$\Delta\varepsilon/kT$  is comparable to $1$, which makes the continuous distribution become step functions.
Comparing the canonical and finite distributions, we can see that the finite distribution is still sharper and has more particles in the condensate state at the zeroth energy level.

\section{CONCLUSIONS AND DISCUSSIONS}\label{sec5}
Motivated by the asynchronous finite differences method proposed in Ref.~\cite{Liu_2022}, which calculates the most probable distributions of finite particle number systems, we checked the numerical method and found that the numerical variation can be used. The central difference can give a more precise estimation of the most probable distributions, and the Stirling approximation can also be removed. The central difference seems to be a natural operation without any artificial manner. We then derived three new finite distributions using the numerical variation and central difference. Although these distributions recover the canonical distributions when the particle number is infinitely large, the new distributions may be significant for finite particle number systems, such as matter in cold atoms, levitation, and Bose-Einstein condensation experiments. Thus, we applied the finite distributions to Bose-Einstein condensation experiments.

By comparing the numerical results of the canonical and finite distributions, we found three important points that could verify the finite distributions
\begin{enumerate}
    \item The critical condition $n\lambda^3 =2.612$
    can be relaxed in finite number systems. If the density is maintained as a constant, a higher critical temperature is expected compared to the infinite number system.
    \item The velocity distributions become sharper than those in the canonical distribution. The number of particles in the condensate state at the zeroth energy level is larger than the prediction from the canonical distribution. More precise measurements on Bose-Einstein condensation could verify this prediction.
    \item When the trapped matter is much smaller, the velocity distributions deviate from the continuous distribution functions if the energy interval is comparable to $kT$.
\end{enumerate}
Our study in this paper 
may give some hints on the frontier of condensed matter
physics, and its application on the real matter and experiments
will be shown in our future work.

\section*{Acknowledgements}
We thank Qiu-Mei Huang, Bo-Yang Liu, Ji-Heng Guo and Yi-Yi Niu 
for their helpful discussions on the numerical analysis. 
This work was supported by the Natural Science Foundation of China under 
grant number 11775012.
\bibliography{refs}

\begin{thebibliography}{14}
\expandafter\ifx\csname natexlab\endcsname\relax\def\natexlab#1{#1}\fi
\expandafter\ifx\csname bibnamefont\endcsname\relax
  \def\bibnamefont#1{#1}\fi
\expandafter\ifx\csname bibfnamefont\endcsname\relax
  \def\bibfnamefont#1{#1}\fi
\expandafter\ifx\csname citenamefont\endcsname\relax
  \def\citenamefont#1{#1}\fi
\expandafter\ifx\csname url\endcsname\relax
  \def\url#1{\texttt{#1}}\fi
\expandafter\ifx\csname urlprefix\endcsname\relax\def\urlprefix{URL }\fi
\providecommand{\bibinfo}[2]{#2}
\providecommand{\eprint}[2][]{\url{#2}}

\bibitem[{\citenamefont{Gonzalez-Ballestero
  et~al.}(2021)\citenamefont{Gonzalez-Ballestero, Aspelmeyer, Novotny, Quidant,
  and Romero-Isart}}]{gonzalez2021levitodynamics}
\bibinfo{author}{\bibfnamefont{C.}~\bibnamefont{Gonzalez-Ballestero}},
  \bibinfo{author}{\bibfnamefont{M.}~\bibnamefont{Aspelmeyer}},
  \bibinfo{author}{\bibfnamefont{L.}~\bibnamefont{Novotny}},
  \bibinfo{author}{\bibfnamefont{R.}~\bibnamefont{Quidant}}, \bibnamefont{and}
  \bibinfo{author}{\bibfnamefont{O.}~\bibnamefont{Romero-Isart}},
  \bibinfo{journal}{Science} \textbf{\bibinfo{volume}{374}},
  \bibinfo{pages}{eabg3027} (\bibinfo{year}{2021}).

\bibitem[{\citenamefont{Jain et~al.}(2016)\citenamefont{Jain, Gieseler, Moritz,
  Dellago, Quidant, and Novotny}}]{jain2016direct}
\bibinfo{author}{\bibfnamefont{V.}~\bibnamefont{Jain}},
  \bibinfo{author}{\bibfnamefont{J.}~\bibnamefont{Gieseler}},
  \bibinfo{author}{\bibfnamefont{C.}~\bibnamefont{Moritz}},
  \bibinfo{author}{\bibfnamefont{C.}~\bibnamefont{Dellago}},
  \bibinfo{author}{\bibfnamefont{R.}~\bibnamefont{Quidant}}, \bibnamefont{and}
  \bibinfo{author}{\bibfnamefont{L.}~\bibnamefont{Novotny}},
  \bibinfo{journal}{Physical review letters} \textbf{\bibinfo{volume}{116}},
  \bibinfo{pages}{243601} (\bibinfo{year}{2016}).

\bibitem[{\citenamefont{Gieseler and Millen}(2018)}]{2018Levitated}
\bibinfo{author}{\bibfnamefont{J.}~\bibnamefont{Gieseler}} \bibnamefont{and}
  \bibinfo{author}{\bibfnamefont{J.}~\bibnamefont{Millen}},
  \bibinfo{journal}{Entropy} \textbf{\bibinfo{volume}{20}},
  \bibinfo{pages}{326} (\bibinfo{year}{2018}).

\bibitem[{\citenamefont{Andrews et~al.}(1997)\citenamefont{Andrews, Townsend,
  Miesner, Durfee, Kurn, and Ketterle}}]{andrews1997observation}
\bibinfo{author}{\bibfnamefont{M.}~\bibnamefont{Andrews}},
  \bibinfo{author}{\bibfnamefont{C.}~\bibnamefont{Townsend}},
  \bibinfo{author}{\bibfnamefont{H.-J.} \bibnamefont{Miesner}},
  \bibinfo{author}{\bibfnamefont{D.}~\bibnamefont{Durfee}},
  \bibinfo{author}{\bibfnamefont{D.}~\bibnamefont{Kurn}}, \bibnamefont{and}
  \bibinfo{author}{\bibfnamefont{W.}~\bibnamefont{Ketterle}},
  \bibinfo{journal}{Science} \textbf{\bibinfo{volume}{275}},
  \bibinfo{pages}{637} (\bibinfo{year}{1997}).

\bibitem[{\citenamefont{Anderson et~al.}(1995)\citenamefont{Anderson, Ensher,
  Matthews, Wieman, and Cornell}}]{anderson1995observation}
\bibinfo{author}{\bibfnamefont{M.~H.} \bibnamefont{Anderson}},
  \bibinfo{author}{\bibfnamefont{J.~R.} \bibnamefont{Ensher}},
  \bibinfo{author}{\bibfnamefont{M.~R.} \bibnamefont{Matthews}},
  \bibinfo{author}{\bibfnamefont{C.~E.} \bibnamefont{Wieman}},
  \bibnamefont{and} \bibinfo{author}{\bibfnamefont{E.~A.}
  \bibnamefont{Cornell}}, \bibinfo{journal}{science}
  \textbf{\bibinfo{volume}{269}}, \bibinfo{pages}{198} (\bibinfo{year}{1995}).

\bibitem[{\citenamefont{Davis et~al.}(1995)\citenamefont{Davis, Mewes, Andrews,
  van Druten, Durfee, Kurn, and Ketterle}}]{davis1995bose}
\bibinfo{author}{\bibfnamefont{K.~B.} \bibnamefont{Davis}},
  \bibinfo{author}{\bibfnamefont{M.-O.} \bibnamefont{Mewes}},
  \bibinfo{author}{\bibfnamefont{M.~R.} \bibnamefont{Andrews}},
  \bibinfo{author}{\bibfnamefont{N.~J.} \bibnamefont{van Druten}},
  \bibinfo{author}{\bibfnamefont{D.~S.} \bibnamefont{Durfee}},
  \bibinfo{author}{\bibfnamefont{D.}~\bibnamefont{Kurn}}, \bibnamefont{and}
  \bibinfo{author}{\bibfnamefont{W.}~\bibnamefont{Ketterle}},
  \bibinfo{journal}{Physical review letters} \textbf{\bibinfo{volume}{75}},
  \bibinfo{pages}{3969} (\bibinfo{year}{1995}).

\bibitem[{\citenamefont{Bradley et~al.}(1995)\citenamefont{Bradley, Sackett,
  Tollett, and Hulet}}]{bradley1995evidence}
\bibinfo{author}{\bibfnamefont{C.~C.} \bibnamefont{Bradley}},
  \bibinfo{author}{\bibfnamefont{C.}~\bibnamefont{Sackett}},
  \bibinfo{author}{\bibfnamefont{J.}~\bibnamefont{Tollett}}, \bibnamefont{and}
  \bibinfo{author}{\bibfnamefont{R.~G.} \bibnamefont{Hulet}},
  \bibinfo{journal}{Physical review letters} \textbf{\bibinfo{volume}{75}},
  \bibinfo{pages}{1687} (\bibinfo{year}{1995}).

\bibitem[{\citenamefont{Liu}(2022)}]{Liu_2022}
\bibinfo{author}{\bibfnamefont{Q.}~\bibnamefont{Liu}}, \bibinfo{journal}{Annals
  of Physics} \textbf{\bibinfo{volume}{441}}, \bibinfo{pages}{168884}
  (\bibinfo{year}{2022}),
  \urlprefix\url{https://doi.org/10.1016%2Fj.aop.2022.168884}.

\bibitem[{\citenamefont{Lao and Zhao}(2021)}]{Lao2021}
\bibinfo{author}{\bibfnamefont{D.}~\bibnamefont{Lao}} \bibnamefont{and}
  \bibinfo{author}{\bibfnamefont{S.}~\bibnamefont{Zhao}},
  \emph{\bibinfo{title}{Direct Methods of Variational Problems}}
  (\bibinfo{publisher}{Springer Singapore}, \bibinfo{address}{Singapore},
  \bibinfo{year}{2021}), pp. \bibinfo{pages}{457--461}, ISBN
  \bibinfo{isbn}{978-981-15-6070-5},
  \urlprefix\url{https://doi.org/10.1007/978-981-15-6070-5_8}.

\bibitem[{\citenamefont{Hanc}(2004)}]{hanc2004original}
\bibinfo{author}{\bibfnamefont{J.}~\bibnamefont{Hanc}},
  \bibinfo{journal}{European Journal of Physics, Submitted}
  p.~\bibinfo{pages}{1} (\bibinfo{year}{2004}).

\bibitem[{\citenamefont{Stahel}(2003)}]{2003Calculus}
\bibinfo{author}{\bibfnamefont{A.}~\bibnamefont{Stahel}},
  \bibinfo{journal}{lecture notes used at hta biel}  (\bibinfo{year}{2003}).

\bibitem[{\citenamefont{Li et~al.}(2017)\citenamefont{Li, Qiao, and
  Tang}}]{nsde}
\bibinfo{author}{\bibfnamefont{Z.}~\bibnamefont{Li}},
  \bibinfo{author}{\bibfnamefont{Z.}~\bibnamefont{Qiao}}, \bibnamefont{and}
  \bibinfo{author}{\bibfnamefont{T.}~\bibnamefont{Tang}},
  \emph{\bibinfo{title}{Numerical solution of differential equations:
  introduction to finite difference and finite element methods}}
  (\bibinfo{publisher}{Cambridge University Press}, \bibinfo{year}{2017}), ISBN
  \bibinfo{isbn}{978-110-71-6322-5}.

\bibitem[{\citenamefont{Greiner et~al.}(2012)\citenamefont{Greiner, Neise, and
  St{\"o}cker}}]{greiner2012thermodynamics}
\bibinfo{author}{\bibfnamefont{W.}~\bibnamefont{Greiner}},
  \bibinfo{author}{\bibfnamefont{L.}~\bibnamefont{Neise}}, \bibnamefont{and}
  \bibinfo{author}{\bibfnamefont{H.}~\bibnamefont{St{\"o}cker}},
  \emph{\bibinfo{title}{Thermodynamics and statistical mechanics}}
  (\bibinfo{publisher}{Springer Science \& Business Media},
  \bibinfo{year}{2012}).

\bibitem[{\citenamefont{Huang}(2008)}]{huang2008statistical}
\bibinfo{author}{\bibfnamefont{K.}~\bibnamefont{Huang}},
  \emph{\bibinfo{title}{Statistical mechanics}} (\bibinfo{publisher}{John Wiley
  \& Sons}, \bibinfo{year}{2008}).

\end{thebibliography}
\end{document}